\documentclass[conference,10pt]{IEEEtran}
\IEEEoverridecommandlockouts
\usepackage[T1]{fontenc}
\usepackage{newtxtext,newtxmath}
\usepackage{amsmath}
\usepackage{graphicx,booktabs,array}
\usepackage{algpseudocode}
\usepackage{float}
\usepackage{cite,flushend,stfloats}
\usepackage[hidelinks]{hyperref}
\usepackage{orcidlink}
\hypersetup{pdftitle={Budgeted Decision-Model-Guided Joint-Action Search for Satellite Edge Computing},pdfsubject={Joint offloading and resource allocation: allocation quality and solver cost},pdfauthor={Ziming Liu, Minghao Sun, Ran Tao, Xiaoli Chu}}
\usepackage{url}
\newcounter{algorithm}
\newenvironment{pseudocode}[1]{%
  \par\addvspace{\medskipamount}\noindent\begin{minipage}{\columnwidth}%
  \small\refstepcounter{algorithm}%
  \hrule height .5pt\kern2pt
  \noindent\textbf{Algorithm~\thealgorithm.} #1\par
  \kern2pt\hrule height .3pt\kern2pt
}{%
  \par\kern2pt\hrule height .5pt
  \end{minipage}\par\addvspace{\medskipamount}%
}
\graphicspath{{figures/}}
\newcommand{\F}{F_{\omega}}

\DeclareMathOperator*{\argmin}{arg\,min}
\DeclareMathOperator*{\argmax}{arg\,max}
\title{Budgeted Decision-Model-Guided Joint-Action\newline Search for Satellite Edge Computing}
\author{\IEEEauthorblockN{\href{https://orcid.org/0009-0002-6772-0005}{Ziming Liu}\,\orcidlink{0009-0002-6772-0005}, \href{https://orcid.org/0009-0009-9936-8828}{Minghao Sun}\,\orcidlink{0009-0009-9936-8828}, \href{https://orcid.org/0000-0002-5176-1948}{Ran Tao}\,\orcidlink{0000-0002-5176-1948}, and \href{https://orcid.org/0000-0003-1863-6149}{Xiaoli Chu}\,\orcidlink{0000-0003-1863-6149}}
\IEEEauthorblockA{School of Electrical and Electronic Engineering, The University of Sheffield, Sheffield, U.K.\\
\href{mailto:ziming.liu@sheffield.ac.uk}{ziming.liu@sheffield.ac.uk}, \href{mailto:msun39@sheffield.ac.uk}{msun39@sheffield.ac.uk}, \href{mailto:ran.tao@sheffield.ac.uk}{ran.tao@sheffield.ac.uk}, \href{mailto:x.chu@sheffield.ac.uk}{x.chu@sheffield.ac.uk}}
\thanks{This work was supported in part by the UK EPSRC grant EP/X038971/1 and the Horizon Europe grants 101086219 and 101212141. Research supported by the NVIDIA Academic Grant Program using NVIDIA-provided RTX PRO 6000 Blackwell Max-Q Workstation Edition GPUs. \newline This work has been submitted to the IEEE for possible publication. Copyright may be transferred without notice, after which this version may no longer be accessible.}}
\usepackage{xcolor}
\begin{document}
\bstctlcite{IEEE3GPPidentity}
\typeout{M13-PAPER: \the\paperwidth, \the\paperheight}
\typeout{M13-TEXT: \the\textwidth, \the\textheight, column=\the\columnwidth}
\typeout{M13-MARGINS: left=\the\oddsidemargin, top=\the\topmargin, sep=\the\columnsep}
\typeout{M13-BODY: baseline=\the\baselineskip}
\maketitle
\begin{abstract}
Efficient task offloading in satellite edge computing requires coordinated allocation of communication and computing resources. Search methods based on large language models (LLMs) have been shown to generate high-quality resource allocations, but the latency of individual model requests and repeated model interactions can lead to substantial solver overhead. We propose budgeted joint-action search (BJAS), an integrated decision-model-guided solver that jointly optimizes satellite offloading and power allocation to minimize average task-completion latency under explicit search budgets. Its bounded candidate pools combine single-terminal updates with power-inheriting offload replacements for batched selection, retaining two-terminal coordination without enumerating additional power vectors. Across 60 completed runs per method under the workflows' respective configurations, BJAS reduces mean total solver time by 97.16\% relative to LLM-generated offloading while achieving a mean task-completion latency close to that of the LLM baseline. BJAS also achieves lower mean task-completion latency than the tested genetic algorithm and deep deterministic policy gradient baselines.
\end{abstract}

\section{Introduction}
\begin{figure}[!t]
\centering\includegraphics[width=\columnwidth]{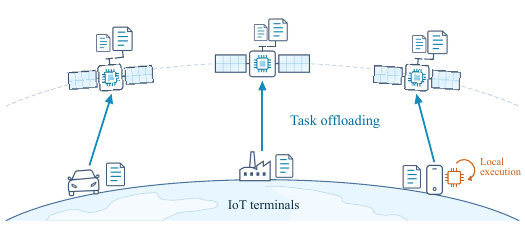}
\caption{Satellite edge computing: IoT terminals execute tasks locally or offload them to satellites with onboard computing resources.}
\label{fig:overview}
\end{figure}

Internet of Things (IoT) terminals may have limited local computing capacity and little access to terrestrial edge servers. Satellite edge computing enables these terminals to offload tasks to satellites~\cite{song2021terrestrial,mao} (Fig.~\ref{fig:overview}). Subchannel reuse and transmit powers determine uplink interference, while offloading assignments determine how satellite computing capacity is shared. An update for one terminal can therefore affect others' completion latencies, motivating joint optimization of offloading decisions, subchannel allocation, transmit powers, and computing-resource allocation~\cite{ding2022joint} to minimize average task-completion latency.

Large language models (LLMs) generate solutions and evolve heuristics from feedback~\cite{opro,reevo}, with applications in wireless planning~\cite{qiu2024wireless} and satellite offloading~\cite{sun}. Repeated model interactions add solver overhead, and model latency is a reported deployment challenge in few-shot wireless resource allocation~\cite{lee2026llmra}. Solver time is the time required to obtain an allocation; task-completion latency comprises the upload and computation delays under that allocation. We aim to reduce solver overhead while retaining allocation quality close to that achieved by LLM-generated offloading. Decision models such as Jev select among supplied candidate updates~\cite{jev} rather than generate complete allocations.

Candidate construction must capture useful terminal interactions within bounded pools. Existing studies consider finite-candidate selection~\cite{demir2026finitepool,yao2025llmopt} and satellite assignment exchanges~\cite{wang2025pricing}. A search restricted to single-terminal updates can miss an improving joint update when each constituent change increases latency (Table~\ref{tab:joint}), whereas enumerating all joint updates can produce large option sets.

Budgeted joint-action search (BJAS) constructs bounded candidate pools for joint satellite offloading and power allocation, uses Jev for batched selection, and numerically verifies improvement before accepting an update. A power-inheriting offload replacement returns an offloaded terminal to local execution while a local terminal starts offloading with the departing terminal's transmit-power vector.

\begingroup
\setlength{\parskip}{0pt}
The main contributions are as follows.
\begin{enumerate}
\item We propose BJAS, which combines single-terminal updates with power-inheriting offload replacements in bounded candidate pools to retain two-terminal coordination without enumerating additional power vectors.
\item Across paired scenarios under the workflows' respective configurations, the complete Jev-based BJAS solver reduces mean total solver time by 97.16\% relative to LLM-generated offloading while achieving a mean task-completion latency close to that of the LLM baseline. BJAS also achieves lower mean task-completion latency than the tested genetic algorithm (GA) baseline and deep deterministic policy gradient (DDPG) adaptation.
\end{enumerate}

\par\endgroup

\section{System Model and Problem Formulation}

\subsection{Network and Channel Models}

\begin{figure}[H]
\centering\includegraphics[width=\columnwidth]{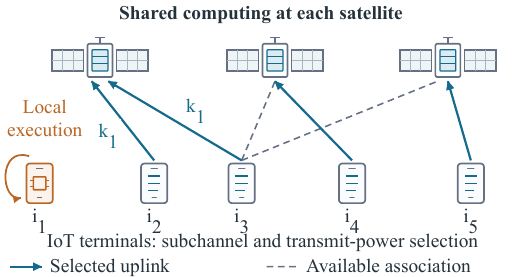}
\caption{Illustrative satellite-edge network with local execution and uplink offloading. Solid links show selected uplinks; dashed links show alternatives for $i_3$. Subchannel $k_1$ is reused; each satellite's computing capacity is shared among assigned tasks.}
\label{fig:concept}
\end{figure}

We consider $M$ satellites, $N$ IoT terminals, and $K$ uplink subchannels, indexed by $\mathcal{M}=\{1,\ldots,M\}$, $\mathcal{N}=\{1,\ldots,N\}$, and $\mathcal{K}_u=\{1,\ldots,K\}$, respectively (Fig.~\ref{fig:concept}). Each terminal $i\in\mathcal{N}$ generates an indivisible task of $D_i$ bits requiring $C_i=cD_i$ cycles, with $c$ measured in cycles/bit. The task executes locally or is offloaded in full to one satellite.

A scenario $\omega$ fixes the task sizes, terminal and satellite positions, and channel realizations throughout optimization and the modeled task batch. A central controller has task and channel state information without estimation or signaling errors. Each satellite uses one fixed receive beam toward the ground-region center; terminals transmit omnidirectionally. Table~\ref{tab:parameters} gives the main-scenario parameters, with carrier, altitude, and antenna hardware informed by the S-band LEO-600 reference in~\cite{tr38821}.

\suppressfloats[t]
\begin{table}[t]
\caption{Main-scenario parameters and evaluation settings}
\label{tab:parameters}\centering\small
\begin{tabular}{@{}>{\raggedright\arraybackslash}p{.56\columnwidth}>{\raggedright\arraybackslash}p{.37\columnwidth}@{}}\toprule
Parameter & Value \\ \midrule
Satellites / terminals / subchannels & $3/10/4$ \\
Task size; intensity $c$ & $0.3$--$0.6$~MB; $100$~cycles/bit \\
Local / satellite capacity $f_d/f_s$ & $0.5/20$~Gcycles/s \\
Per-terminal total power & $P_{\max}=200$~mW \\
Carrier $f_c$; total bandwidth $KB$ & $2$~GHz; $40$~MHz \\
Subchannel bandwidth $B$ & $10$~MHz for $K=4$ \\
Altitude; minimum service elevation & $600$~km; $10^\circ$ \\
Earth / ground-region radius & $6371$~km / $20$~km \\
Terminal / peak receive gain & $0/30$~dBi \\
Receive aperture diameter $D_{\mathrm a}$ & $2$~m \\
Receiver $G/T$; temperature $T_{\mathrm{sys}}$ & $1.1$~dB/K; $776.247$~K \\
Rician factor & $\kappa=10$ (linear) \\
Noise power $\sigma^2$ per subchannel & $1.071725\!\times\!10^{-10}$~mW \\
Search seeds & $101,202,303$ \\
Scenarios $\times$ repeats & $20\times3$ \\
\bottomrule
\end{tabular}
\end{table}

On a spherical Earth of radius $r_e$, terminals are sampled uniformly by surface area within a geodesic radius of $20$~km around $\mathbf q_0=(0,0,r_e)^{\mathsf T}$. Satellites lie at altitude $h$, with elevations viewed from $\mathbf q_0$ and azimuths independently sampled from $\mathcal U[10^\circ,90^\circ]$ and $\mathcal U[0,2\pi)$, respectively. Let $\mathbf q_i$ and $\mathbf s_m$ denote terminal and satellite positions, with $\|\mathbf q_i\|=r_e$ and $\|\mathbf s_m\|=r_e+h$. The link distance and elevation $\varepsilon_{mi}\in[-\pi/2,\pi/2]$ are
\begin{align}
d_{mi}&=\|\mathbf s_m-\mathbf q_i\|,\nonumber\\
\sin\varepsilon_{mi}
&=\frac{(\mathbf s_m-\mathbf q_i)^{\mathsf T}\mathbf q_i}{d_{mi}r_e}.
\label{eq:distance}
\end{align}

The fixed boresight and off-axis angle $\theta_{mi}\in[0,\pi]$ are defined by
\begin{equation}
\mathbf b_m=\frac{\mathbf q_0-\mathbf s_m}{\|\mathbf q_0-\mathbf s_m\|},
\quad
\cos\theta_{mi}=\mathbf b_m^{\mathsf T}\frac{\mathbf q_i-\mathbf s_m}{d_{mi}}.
\label{eq:boresight}
\end{equation}
For $u(\theta)=\pi D_{\mathrm a}\sin\theta/\lambda$, where $\lambda=c_0/f_c$, the normalized circular-aperture power pattern is~\cite[Sec.~6.4.1]{tr38811}
\begin{equation}
A(\theta)=
\begin{cases}
1,&\theta=0,\\
\left[2J_1(u(\theta))/u(\theta)\right]^2,&0<\theta\leq\pi/2,\\
0,&\pi/2<\theta\leq\pi.
\end{cases}
\label{eq:pattern}
\end{equation}
Here, $J_1$ is the first-order Bessel function of the first kind and $c_0$ is the speed of light. The model assumes zero back-hemisphere gain and specifies peak receive gain separately.

The indicators $v_{mi}=\mathbf1\{\varepsilon_{mi}>0\}$ and $\eta_{mi}=\mathbf1\{\varepsilon_{mi}\geq10^\circ\}$ denote physical visibility and service eligibility, respectively. A visible link below the service threshold can contribute interference even though it cannot be selected for offloading. Using free-space propagation and a normalized scalar Rician model~\cite[Secs.~6.6.2 and~6.9.2]{tr38811}, the channel-power gain is
\begin{align}
G_{mik}
&=v_{mi}g_{\mathrm t}g_{\mathrm r,0}A(\theta_{mi})
\left(\frac{\lambda}{4\pi d_{mi}}\right)^2
|h_{mik}|^2,\nonumber\\
h_{mik}
&=\sqrt{\frac{\kappa}{\kappa+1}}
+\frac{\xi_{mik}+\mathrm{j}\zeta_{mik}}
{\sqrt{2(\kappa+1)}},
\label{eq:gain}
\end{align}
where $g_{\mathrm t}=1$ and $g_{\mathrm r,0}=1000$ are linear power gains, and $\xi_{mik},\zeta_{mik}$ are independent standard Gaussian samples across links and subchannels, giving $\mathbb E[|h_{mik}|^2]=1$. A link's subchannels share geometry and antenna gains; additional propagation losses are neglected and polarization is ideally matched.

Let $\Gamma$ denote the receiver $G/T$ in dB/K. The system noise temperature in kelvin and noise power per subchannel in milliwatts follow from the link-budget relation in~\cite[Sec.~6.1.3.1]{tr38821}:
\begin{align}
T_{\mathrm{sys}}&=\frac{g_{\mathrm r,0}}{10^{\Gamma/10}},\nonumber\\
\sigma^2&=10^3k_{\mathrm B}T_{\mathrm{sys}}B,
\label{eq:noise}
\end{align}
where $k_{\mathrm B}=1.380649\times10^{-23}$~J/K. The system noise temperature is derived from peak receive gain and remains fixed across links; all subchannels have bandwidth $B$.

\subsection{Offloading, Uplink Transmission, and Computing Latency}

Let $z=[z_{mi}]\in\{0,1\}^{M\times N}$, with $z_{mi}=1$ indicating offloading from terminal $i$ to satellite $m$. The constraints are $o_i=\sum_{m=1}^{M}z_{mi}\leq1$ and $z_{mi}\leq\eta_{mi}$; $o_i=0$ denotes local execution.

Let $P=[p_{ik}]\in\mathbb{R}_{\geq0}^{N\times K}$, where $p_{ik}$ is terminal $i$'s power on subchannel $k$ in mW, with $\sum_{k=1}^{K}p_{ik}\leq P_{\max}o_i$. Local terminals therefore have zero power; offloaded terminals require positive total power and uplink rate. Terminals may reuse subchannels and transmit over multiple subchannels. The allocation determines the number of offloaded tasks.

For an offloaded terminal $i$, let $m(i)$ denote its serving satellite and $I_{mik}$ the interference at satellite $m$ on subchannel $k$, excluding terminal $i$. Using the link budget in~\cite[Sec.~6.1.3.1]{tr38821}, we define receiver-specific interference and adopt the idealized Shannon-form uplink rate:
\begin{align}
I_{mik}
&=\sum_{\substack{j=1\\j\neq i}}^{N}
o_jp_{jk}G_{mjk},
\nonumber\\
R_i
&=B\sum_{k=1}^{K}
\log_2\!\left(
1+\frac{p_{ik}G_{m(i)ik}}
{\sigma^2+I_{m(i)ik}}
\right),
\label{eq:rate}
\end{align}
Here, $G_{mik}$ is dimensionless, $B$ is in Hz, $I_{mik}$ is in mW, and $R_i$ is in bit/s. Each transmitter contributes interference through its channel $G_{mjk}$ to the target receiver, regardless of its serving satellite. Subchannels are orthogonal and frequency-flat; co-channel interference is treated as noise under single-user decoding. Equation~\eqref{eq:rate} assumes independent zero-mean Gaussian terminal signals, additive white Gaussian receiver noise, and ideal coding; protocol overhead is excluded. All uploads start simultaneously at rates set by the initial active set. Each transmitter retains its rate and powers until its upload completes, then stops; remaining terminals retain their selected rates. Timing synchronization and frequency-offset compensation are ideal; residual Doppler and inter-subchannel leakage are neglected.

Define $\mathcal{L}=\{i\in\mathcal{N}:o_i=0\}$ and $\mathcal{S}_m=\{i\in\mathcal{N}:z_{mi}=1\}$. Each terminal has local capacity $f_d$ and each satellite has total capacity $f_s$, in cycles/s. Offloaded tasks begin computing after upload with fixed CPU reservations, which are not redistributed when tasks finish. For $i\in\mathcal{S}_m$, let $f_{mi}>0$ be the reserved computing rate in cycles/s, with $\sum_{i\in\mathcal{S}_m}f_{mi}\leq f_s$.

For a fixed assignment, minimizing aggregate computing delay $\sum_{i\in\mathcal{S}_m}C_i/f_{mi}$ gives~\cite{sun}
\begin{equation}
f_{mi}^{\star}
=\frac{f_s\sqrt{C_i}}
{\sum_{j\in\mathcal{S}_m}\sqrt{C_j}},
\qquad i\in\mathcal{S}_m,
\label{eq:cpu}
\end{equation}
which yields the minimum aggregate computing delay
\begin{equation}
\sum_{i\in\mathcal{S}_m}\frac{C_i}{f_{mi}^{\star}}
=\frac{\left(\sum_{i\in\mathcal{S}_m}\sqrt{C_i}\right)^2}{f_s}.
\label{eq:cpu_delay}
\end{equation}
An empty satellite task set contributes zero delay. Local and offloaded tasks take $C_i/f_d$ and $D_i/R_i+C_i/f_{mi}^{\star}$, respectively. Substituting the optimal computing rates gives the average task-completion latency objective over $x=(z,P)$:
\begin{equation}
\begin{split}
\min_{x\in\mathcal{X}_\omega}\quad
F_\omega(x)=\frac{1000}{N}\Bigg[
&\sum_{i\in\mathcal{L}}\frac{C_i}{f_d}
+\sum_{i\notin\mathcal{L}}\frac{D_i}{R_i}\\
&+\sum_{m=1}^{M}
\frac{\left(\sum_{i\in\mathcal{S}_m}\sqrt{C_i}\right)^2}{f_s}
\Bigg],
\end{split}
\label{eq:objective}
\end{equation}

Here, $F_\omega(x)$ is average task-completion latency in ms for scenario $\omega$, and $\mathcal{X}_\omega$ enforces the association, service eligibility, power, and positive-rate constraints.

Task sizes are independently sampled from a uniform distribution over $[0.3,0.6]$~MB, where $1$~MB corresponds to $8\times10^6$ bits. The objective excludes propagation, queueing, and result-return delays.

\subsection{Task-Completion Latency and Solver Cost}
\label{sec:solver-cost}

Allocation quality is measured by $\F$. Solver cost comprises client-side model API attempts $H$, full objective evaluations $E$, and wall-clock time $T_{\mathrm{solve}}$. $H$ includes failures and retries; a request selecting from several pools counts once. A full objective evaluation computes $F_\omega(x)$ for one allocation, including analytic CPU allocation. $E$ includes initialization evaluations and excludes cache hits. Solver time is measured from preparation to the final checkpoint and includes startup, candidate preparation, client-observed model requests, numerical evaluation, and recording overhead. For quality--time curves, an allocation becomes available only after evaluation, validation, and saving.

\begingroup
\setlength{\intextsep}{4pt plus 2pt minus 2pt}
\begin{table}[!t]
\caption{Task-completion latency under individual and joint updates. Reductions are relative to the current allocation.}
\label{tab:joint}\centering\small
\setlength{\tabcolsep}{3pt}
\begin{tabular*}{\columnwidth}{@{\extracolsep{\fill}}lrrrr@{}}\toprule
 & Incumbent & Local 9 & Offload 7 & Joint\\\midrule
Latency (ms) &578.417&599.714&604.979&577.599\\
Reduction (ms) &---&$-21.297$&$-26.562$&$0.818$\\\bottomrule
\end{tabular*}
\end{table}
\endgroup

\begin{table}[!b]
\caption{Mean and best task-completion latency across 20 network scenarios with three runs per method and scenario.}
\label{tab:quality}\centering\small
\renewcommand{\arraystretch}{1.12}
\begin{tabular*}{\columnwidth}{@{\extracolsep{\fill}}lrr@{}}\toprule
Solver & Mean (ms) & Best (ms)\\\midrule
\textbf{BJAS} & \textbf{618.404} & \textbf{519.563}\\
LLM & 618.075 & 519.873\\
GA & 639.557 & 537.183\\
DDPG & 671.245 & 571.282\\
\bottomrule
\end{tabular*}
\end{table}

\section{Decision-Model-Guided Joint-Action Search}
\subsection{Bounded Pools and Coordinated Candidate Updates}
\label{sec:candidates}
BJAS initializes by evaluating the all-local allocation $x_{\rm local}$ once (Algorithm~\ref{alg:bjas}). At iteration $t$, BJAS cyclically selects a set $\mathcal W_t$ of terminals. For each $j\in\mathcal W_t$, it constructs a pool $\mathcal C_j$ of single-terminal or joint updates involving $j$, all relative to the incumbent $x_t$.

BJAS restricts the continuous powers in~\eqref{eq:objective} to feasible discrete codebook allocations: each offloaded terminal transmits with total power $\rho P_{\max}$, $\rho\in\{0.5,1\}$, on one subchannel or uniformly across all subchannels.

Caps $B_E,B_H,B_R$ limit full evaluations, model requests, and rounds, respectively; elapsed time $\tau$ in seconds is checked against $B_T$ ($B_T=\infty$: unlimited). Let $\mathcal K$ and $\mathcal H$ denote the cache and feedback history, respectively. The allocation $x\oplus a$ results from applying all updates in $a$ to $x$.

\begin{figure}[!t]
\centering{\setlength{\fboxsep}{1.5pt}\setlength{\fboxrule}{0pt}%
\fbox{\includegraphics[width=\dimexpr0.92\columnwidth-2\fboxsep-2\fboxrule\relax]{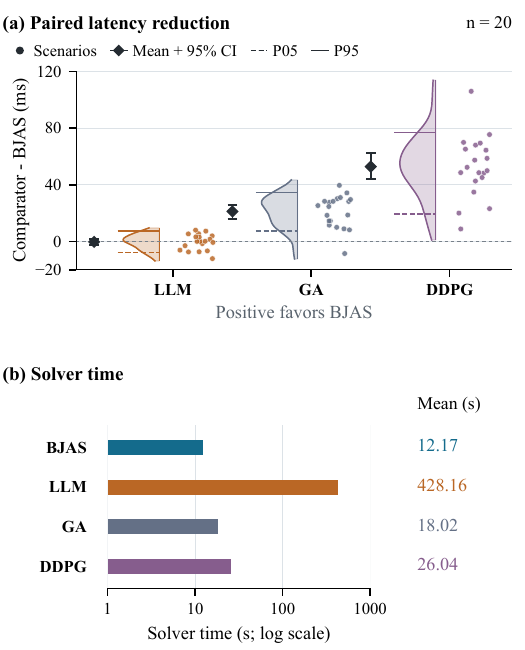}}}
\caption{Allocation quality and solver time across 20 network scenarios with three repeats each. (a) Half-violins show paired latency reductions (comparator minus BJAS); points denote scenario means. Black diamonds and vertical intervals denote means and 95\% confidence intervals from bootstrap resampling of scenarios. Dashed and solid lines indicate the 5th and 95th percentiles, respectively. Positive values favor BJAS. (b) Mean solver time over 60 completed runs per method; logarithmic bars start at 1 s.}
\label{fig:results}
\end{figure}

Candidates include local execution, insertion on idle or reused subchannels, satellite migration, subchannel reassignment, power adjustment, simultaneous changes to a terminal's serving satellite and power vector, and power-inheriting offload replacement. BJAS removes no-ops, duplicates, and cached complete allocations from each pool and groups the remaining candidates by update type. When pool capacity permits, the candidate with the largest link hint is retained for each update type. The link hint sums the subchannel spectral efficiencies of modified terminals that offload under the candidate, after removing all modified terminals' old interference contributions. For a replacement, only the incoming terminal contributes to the hint. The remaining pool slots are filled by stratified sampling across update types, satellites, and active subchannels. Candidates are randomly ordered within each pool; each candidate is linked to its parent allocation and the terminal updates it specifies.

A \emph{power-inheriting offload replacement} returns offloaded terminal $i$ to local execution and offloads local terminal $j$ to an eligible satellite using $i$'s power vector. It preserves the offloaded-task count and total transmit power per subchannel. For a fixed departing terminal, incoming terminal, and target satellite, inheritance reduces the $Q$ distinct nonzero codebook power choices to one ($Q=10$ here).

\subsection{Numerical Context and Structured Candidate Ranking}
\label{sec:ranking}
Context $s$ in Algorithm~\ref{alg:bjas} contains the incumbent, tasks, satellite loads, subchannel signals, and recent feedback. Preprocessing computes upload and computation hints for modified terminals using the resulting interference within and outside the modified set and the final satellite loads. Full numerical evaluation supplies complete candidate objectives, including updated upload delays for unchanged offloaded terminals; preprocessing cost is recorded as preparation time.

Jev (\texttt{jev-1.13.0}) assigns within-pool option probabilities $\pi_j(a)$ from structured text~\cite{jev}. Algorithm~\ref{alg:bjas} selects each pool's highest-probability candidate, directly selecting singleton candidates. Probabilities are compared only within pools; complete numerical objectives determine acceptance across pools.

Up to four pools of at most eight candidates form one batch with a common parent; non-singleton pools share one ranking request~\cite{fanout}. Selected updates are evaluated separately against that parent using the complete objective. Each batch accepts at most one improving update.

\begin{pseudocode}{Budgeted joint-action search (BJAS)}\label{alg:bjas}
\begin{algorithmic}[1]
\Require $\omega$; caps $B_E,B_H,B_R,B_T$; $\epsilon=10^{-9}$ ms
\State $x\gets x_{\rm local}$; $f\gets\F(x)$; $e\gets1$; $n_H,u\gets0$
\State $\mathcal K\gets\{x\mapsto f\}$; $\mathcal H\gets\varnothing$
\For{$t=1,\ldots,B_R$}
  \State \textbf{if} $e\ge B_E$ or $\tau\ge B_T$ \textbf{then break}
  \State $\bar x\gets x$; $\mathcal W\gets\Call{Rotate}{\min\{4,N,B_E-e\}}$
  \State $(\{\mathcal C_j\},s)\gets\Call{PreparePools}{\bar x,\mathcal W,\mathcal K,\mathcal H}$
  \State $\mathcal J\gets\{j:|\mathcal C_j|>0\}$; $\mathcal J_+\gets\{j:|\mathcal C_j|>1\}$
  \State $u\gets(u+|\mathcal W|)\mathbf1_{\{\mathcal J=\varnothing\}}$
  \State \textbf{if} $u\ge N$ or $\tau\ge B_T$ \textbf{then break}
  \State \textbf{if} $\mathcal J=\varnothing$ \textbf{then continue}
  \State $\pi_j(a)\gets1$, $j\in\mathcal J\setminus\mathcal J_+$, $a\in\mathcal C_j$
  \If{$\mathcal J_+\ne\varnothing$}
    \State \textbf{if} $n_H\ge B_H$ \textbf{then break}
    \State $n_H\gets n_H+1$; $\pi_{\mathcal J_+}\gets\Call{RankValidate}{s,\{\mathcal C_j\}_{j\in\mathcal J_+}}$
    \State \textbf{if} $\pi_{\mathcal J_+}=\bot$ \textbf{then break}
  \EndIf
  \State $a_j\gets\argmax_{a\in\mathcal C_j}\pi_j(a)$, $j\in\mathcal J$
  \State $\mathcal Y\gets\langle\bar x\oplus a_j\rangle_{j\in\mathcal J}$
  \State $(\mathcal O,\Delta e)\gets\Call{EvalBatch}{\mathcal Y,\mathcal K,B_E-e,B_T}$
  \State $e\gets e+\Delta e$; $(y,g)\gets\argmin_{(x^{\prime},v)\in\langle(x,f)\rangle\Vert\mathcal O_f}v$
  \State \textbf{if} $g<f-\epsilon$ \textbf{then} $(x,f)\gets(y,g)$
  \State $\mathcal H\gets\Call{Feedback}{\mathcal H,\bar x,\mathcal O,x}$
\EndFor
\State \Return $(x,f)$
\end{algorithmic}
\end{pseudocode}

\begin{figure*}[!t]
\centering{\setlength{\fboxsep}{1.5pt}\setlength{\fboxrule}{0pt}%
\fbox{\includegraphics[width=\dimexpr\textwidth-2\fboxsep-2\fboxrule\relax]{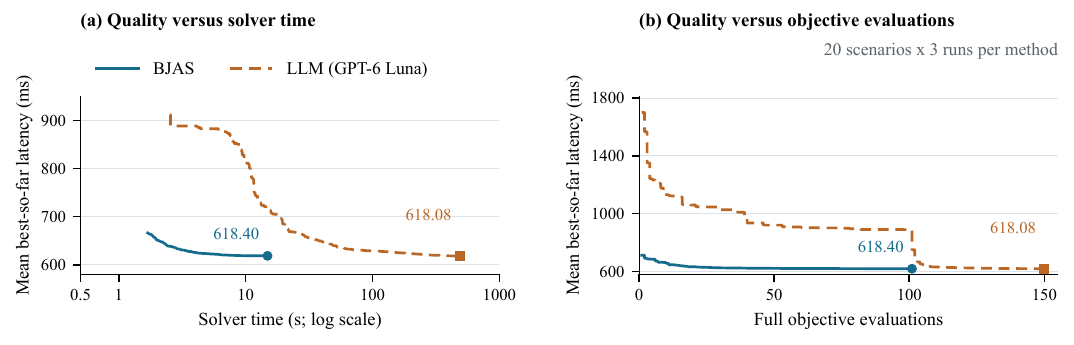}}}
\caption{Mean best-so-far task-completion latency across 20 network scenarios with three runs per method: (a) elapsed solver time based on when validated solutions become available; (b) cumulative full objective evaluations, including initialization. Vertical scales differ between panels.}
\label{fig:convergence}
\end{figure*}

\subsection{Acceptance, Accounting, and Algorithmic Properties}
Algorithm~\ref{alg:bjas} uses $f,e,n_H,\bar x$ for the retained objective, full-evaluation count, request count, and batch parent. Counter $u$ accumulates $|\mathcal W|$ over consecutive empty batches; any nonempty pool resets it. \textsc{RankValidate} returns validated rankings or $\bot$ on request or validation failure. \textsc{EvalBatch} checks budgets before each evaluation, updates $\mathcal K$ in place, and returns ordered outcomes $\mathcal O$ and additional full-evaluation count $\Delta e$. $\mathcal O_f$ is obtained from $\mathcal O$ by retaining, in order, the pairs $(x^{\prime},\F(x^{\prime}))$ for feasible allocations with finite objective values. Angle brackets denote ordered lists, $\Vert$ concatenates them, and ties are broken by selecting the first occurrence.

With feasible initialization of finite objective value and feasibility validation before acceptance, the strict improvement test in Algorithm~\ref{alg:bjas} preserves feasibility and a nonincreasing retained objective.

Cache keys contain the scenario, evaluator version, and complete allocation. A rejected update operation can be reconsidered if it yields an uncached complete allocation when applied to a new parent. Feedback records outcomes and the accepted allocation.

Main experiments use $B_H=25$ model requests, $B_E=501$ full evaluations including initialization, $B_R=1000$ rounds, and $B_T=\infty$.

\subsection{LLM-Generated Offloading Baseline}
In each round, the LLM baseline generates a complete allocation in a single JSON response containing terminal, satellite, and power-vector fields. The allocation is converted to $(z,P)$, validated locally, and evaluated using the same numerical objective.
The search starts from the best of 100 feasible random allocations and uses GPT-6 Luna with medium reasoning effort for up to 50 rounds. The all-local reference is computed separately. Valid proposals are accepted only upon improvement; invalid outputs are recorded and skipped, and service failure terminates the attempt. Automatic and format-repair retries are disabled. Prompts include state, incumbent, stagnation count, and outcome feedback.

\section{Simulation Results}
\subsection{Experimental Setup and Evaluation Protocol}
We compare the complete Jev-based BJAS, LLM-generated offloading, GA, and DDPG workflows using the latency and solver-time definitions in Section~II-C, including initialization under each method's stated settings. For each of the 20 scenarios, all methods use the same task and channel samples, analytic CPU-allocation rule, and numerical objective, with three completed runs per method (seeds 101, 202, and 303). In Table~\ref{tab:quality}, Best denotes the lowest final task-completion latency across the 60 runs for each method. For each method, we first average latency and solver time over the three runs within each scenario and then average the resulting scenario means with equal weights.

\begingroup
\setlength{\intextsep}{2pt plus 1pt minus 1pt}
\begin{table}[H]
\caption{Model-request latency and the proportion of solver time spent on requests}
\label{tab:requests}\centering\small
\begin{tabular}{@{}lrrrrr@{}}\toprule
Solver & Runs & Req./run & Median (s) & P95 (s) & Req. time (\%)\\\midrule
BJAS & 60 & 25 & 0.298 & 0.335 & 62.3\\
LLM & 60 & 50 & 9.100 & 11.919 & 97.5\\
\bottomrule
\end{tabular}
\end{table}
\endgroup

GA uses 400 individuals and 350 generations. DDPG~\cite{ddpg} uses 100 random initialization evaluations and 5000 training/optimization steps; action and policy evaluations are counted separately. DDPG selects the best allocation across initialization, exploration, and policy checks. BJAS and LLM settings follow Section~III.

Experiments are conducted on an AMD Ryzen Threadripper 9960X workstation (24 cores, 192 GiB RAM). GA and DDPG run on the CPU; model requests are served remotely. Independent runs are timed separately.

\begin{samepage}
Paired 95\% percentile-bootstrap confidence intervals use 10000 resamples of the 20 scenarios (seed 8675309)~\cite{agarwal}. Latency reductions are comparator minus BJAS; positive values favor BJAS. Win/tie/loss counts use a $10^{-6}$~ms tolerance.\par
\end{samepage}

\subsection{Allocation Quality and Solver Time}

The complete BJAS workflow reduces mean total solver time by 97.16\% relative to LLM-generated offloading, from 428.162 s to 12.173 s, under the workflows' respective initialization procedures, search budgets, and execution settings (Fig.~\ref{fig:results}(b)). The same 60 completed runs per method yield close observed mean task-completion latencies: 618.40 ms for BJAS and 618.08 ms for the LLM baseline (Table~\ref{tab:quality}). BJAS records the shortest mean solver time among the four methods.

The paired mean LLM-minus-BJAS difference is $-0.329$ ms, with a 95\% bootstrap interval from $-2.696$ to 1.876 ms and 10 wins, two ties, and eight losses (Fig.~\ref{fig:results}(a)). Relative to GA and DDPG, BJAS reduces mean latency by 21.152 and 52.840 ms, with 95\% bootstrap intervals of [15.978, 25.887] and [43.813, 62.345] ms, respectively. The corresponding win/tie/loss counts are 19/0/1 and 20/0/0.

Table~\ref{tab:joint} illustrates the role of coordination using individual and joint updates from the same incumbent in scenario 1 with search seed 101. Terminal 9 returns to local execution, and terminal 7 offloads to satellite 3 on subchannel 1 with the inherited 200 mW power. The joint replacement reduces mean task-completion latency by 0.818 ms, while the two constituent changes individually increase it by 21.297 and 26.562 ms, respectively.

\subsection{Recorded Model-Interaction Overhead}
Client-observed request latency is the elapsed time for each model request, including transport, service waiting, and response handling. Table~\ref{tab:requests} reports request-latency medians and 95th percentiles pooled over requests from each method's completed runs. The request-time share is computed per run as the sum of its nonoverlapping request durations divided by solver time, then averaged over completed runs.

Across completed runs, BJAS spends a mean of 7.536 s on 25 ranking requests per run (up to four pools each), compared with 417.714 s on 50 full-allocation requests for the LLM baseline. Mean BJAS candidate-preparation time is 0.359 s per run. Accumulated model-request time accounts for most of the observed difference in total solver time.

\subsection{Quality Versus Elapsed Solver Time}
Fig.~\ref{fig:convergence}(a) tracks the best-so-far allocations available to each workflow, using matched quality and timing records from the same 60 runs per method as Table~\ref{tab:quality}. Each method's time curve starts when all 60 runs have an available allocation: 1.692 s for BJAS and 2.557 s for the LLM baseline. A completed run's final latency is carried forward in the mean until the last run for that method ends.

At the 10-s comparison checkpoint in Fig.~\ref{fig:convergence}(a), BJAS achieves a 23.85\% lower mean best-so-far task-completion latency than the LLM baseline (618.509 versus 812.263 ms).

Fig.~\ref{fig:convergence}(b) compares allocation quality against full objective evaluations, counting the BJAS all-local evaluation and all 100 LLM random initial evaluations. In the reported runs, BJAS and the LLM baseline use at most 101 and 150 full objective evaluations per run, respectively; the separately computed all-local reference for the LLM baseline is excluded.

\section{Conclusion}
BJAS jointly optimizes satellite offloading and power allocation under explicit search budgets using bounded pools of single-terminal updates and power-inheriting offload replacements.
\newpage
\noindent Power inheritance retains two-terminal coordination without enumerating additional power vectors; Jev selects candidates in batches, and BJAS numerically verifies improvement before accepting an update. Client-side timing records show that accumulated model-request time accounts for most of the observed solver-time difference. With a 25-request budget, BJAS achieves an observed mean task-completion latency close to that of the 50-request LLM baseline while reducing mean total solver time over completed runs by 97.16\% across 20 paired network scenarios under the workflows' respective configurations.

\par

\bibliographystyle{IEEEtran}
\bibliography{references}
\end{document}